\documentstyle[12pt,aasms4]{article}

\def\sun{\hbox{$\odot$}}

\lefthead{}
\righthead{}

\begin{document}
\title{Exploring the Leo II Dwarf Spheroidal: I. The Variable Star Content}

\author{M. H. Siegel and S. R. Majewski\altaffilmark{1}}
\affil{University of Virginia, Department of Astronomy}
\authoraddr{P.O. Box 3018, Charlottesville, VA  22903-0818\\
e:mail: mhs4p@virginia.edu, srm4n@didjeridu.astro.virginia.edu}

\altaffiltext{1}{David and Lucile Packard Foundation Fellow, Cottrell Scholar of the Research
Corporation}

\begin{abstract}
We present the first comprehensive catalogue of variable stars in the Leo II dwarf spheroidal galaxy.  We 
have identified 148 RR Lyrae type variables, of which 140 were amenable to 
derivation of variability parameters with our data.  We have also confirmed the existence
of four anomalous Cepheids as identified in previous studies.

The average period of the RR Lyrae ab 
variables (0.62 days), the fraction of c variables (0.24) and the minimum 
period of the RR Lyrae ab variables (0.51 days) all define Leo II as 
an ``Oosterhoff intermediate" galaxy.  We have used the properties of these
variables to derive a metallicity for Leo II of approximately [Fe/H]=-1.9.  We attempt to resolve 
discrepancies between this value and 
those determined by previous efforts.  The presence of longer period, higher amplitude
RR Lyrae variable implies a metallicity distribution that extends to as poor as [Fe/H]=-2.3.

Leo II's location on the period-metallicity relation of clusters, like that
of other ``Ootershoff intermediate'' objects, falls
between the Oosterhoff Class I and Oosterhoff Class II clusters.
The properties of the variable populations of these objects are consistent with the idea 
that the Oosterhoff ``dichotomy" is a continuum.  The gap between the classes
seems to be explained by the horizontal branch of Galactic globular clusters shifting away from
the instability strip at at intermediate metallicities.
However, Leo II, as well as other Oosterhoff intermediate objects,
has a second parameter effect strong enough to leave
horizontal branch stars in the instability strip.

\end{abstract}

\section{Introduction}

The Leo II dwarf spheroidal galaxy (dSph) was first identified by Harrington and Wilson (1950)
on the Palomar Sky Survey.  Detailed information about Leo II did not appear
in the literature for over three decades due to its extreme distance and 
small size (except the starcount study of Hodge 1962 and abstracts on variables in Swope 1967 and 
Swope 1968) .  
A wave of interest in the 1980's (Hodge 1982; Demers and Harris 1983, hereafter DH83; 
Aaronson et.al. 1983; Azzopardi et al. 1985;
Suntzeff et al. 1986, S86) revealed Leo II to be a metal-poor, second-parameter horizontal branch
object with a handful of carbon stars.

Recently, a series of studies have expanded our knowledge of the Leo II system.  
Demers \& Irwin (1993, DI93) with deep BV CCD photometry,  
derived a metallicity of [Fe/H] = -1.9 and revised Leo II's distance to 215 kpc.  
This was followed by the VI CCD studies of Lee (1995, hereafter L95) and Mighell
\& Rich (1996, hereafter MR96).  The former derived a metallicity of -1.97 while 
the latter derived a metallicity of -1.6 and determined that Leo II formed almost
all of its stars between 7 and 14 Gyr ago.
Both confirmed a distance modulus close to the DI93 value.  
Vogt et al. (1995) measured radial velocities in Leo II
and derived a mass-to-light ratio of 11.1.

One of the gaps in our knowledge of Leo II remains its variable star 
content.  The seminal work in this field was to be the comprehensive survey of Baade 
and Swope.  Their collection of over a hundred Palomar 200-inch plates was intended
to produce 
a complete catalogue of Leo II variables.  Unfortunately, while the catalogue exists, it has not been
published, nor has a rigorous analysis been applied to the data.  Swope did publish two brief 
abstracts, one reporting the identification of 152 variables and 
period measurement for 76 (Swope 1967) and another reporting four anomalous 
Cepheids (Swope 1968).  A later update by van Agt (1973) reported 64 Bailey (1902) type
ab RR Lyrae stars, six type c and an average ab period of 0.59 days.  Neither 
report was considered complete by the authors and neither included 
coordinates, finding charts or light curves.
DI93 reported 
the identification of 80 variable candidates, but did not have enough 
images to fit light curves.

In this paper, we publish the first comprehensive list of positively identified variable 
stars in Leo II.  We have found 148 RR Lyrae variables, for which we have 
successfully derived periods and amplitudes for 140.  The distribution 
of periods is very similar to 
other dSph galaxies, with a moderate fraction of RRc variables.  All of our
observations are consistent with an ``Oosterhoff intermediate'' classification for 
Leo II and this classification is exhibited on a star-by-star basis.  We have
also identified four anomalous Cepheid variables.  The
presence and characteristics of these populations confirm that Leo II is metal-poor
with a large intermediate age population.  The longest period RRab variables hint at the 
existence of Leo II stars as metal-poor as [Fe/H]=-2.3.

\section{Observations and Reduction}

We observed the Leo II dSph with the 4-meter Mayall telescope at Kitt Peak on 
UT 8-9 April 1997 and 22-23 February 1998.  Both observing runs used the T2KB $2048^2$ thinned 
CCD chip and standard Harris prescription UBV filters.  Our first observing run
used the old Mayall prime focus doublet corrector while our second used the new four-element corrector
(now in use with the MOSAIC camera).  To avoid the ghost image produced by the four-element
corrector, the camera was mounted ten minutes off of the optical axis of the telescope.  This increased
the distortion of the images, but not to a level that was uncorrectable.  
All data were reduced by the standard CCDPROC pipeline in 
IRAF.\footnote[1]{IRAF is distributed by the National Optical Astronomy Observatories,
which are operated by the Association of Universities for Research
in Astronomy, Inc., under cooperative agreement with the National
Science Foundation.}

All four nights suffered from very non-photometric observing conditions.  
We are, however, able to transform every frame to an
identical instrumental system by inter-comparing individual photometric measures from frame to frame.
This correction was applied iteratively until the average frame-to-frame residuals were reduced to 0.001
magnitudes.
The combined instrumental magnitudes have been calibrated to the 
BV data set of DI93 within 0.01 magnitudes by the application of zero
point and color corrections.

\begin{center}
$m_V = V_{inst} - 0.29 - 0.12 (B-V) + 0.12 (B-V)^2$

$m_B = B_{inst} - 0.20 - 0.04 (B-V) + 0.10 (B-V)^2$
\end{center}

This transformation does leave some non-linearity in the comparison (Figure 1).  This non-linearity is 
approximately 0.1 magnitudes over the four magnitudes of comparison.  This may be due to 
filter differences or a non-linearity in the CCD
used in one of the two studies.  Independent calibration of the data set will be clearly be 
required in the future.  Because of this non-linearity, all variable measures and light curve 
determinations were made purely from the instrumental system.  However, all average magnitudes
and intensity-weighted mean magnitudes in our catalogues have been converted to the DI93 system via the 
equations listed above.

We found that a total of 72 Leo II UBV images were usable for photometry.  All of these images
were processed through DAOPHOT (Stetson, 1987) and ALLFRAME (Stetson, 1994).  
The {\it combined} photometry
is complete to a depth of $B=V=24.5$ with instrumental errors at the horizontal branch of 
Leo II (V=22.1) of $(\sigma_{V},\sigma_{B})=(0.008,0.016)$.  However, the {\it individual}
images have a wide variation in quality.  Their average completeness limit is $B=V=23.5$
with individual frame errors of $(\sigma_{V},\sigma_{B})=(0.08,0.06)$ at the horizontal
branch.

Accurate stellar positions were derived by using the IRAF task TFINDER with the Hubble Space Telescope 
Guide Star Catalogue.  We were able to derive centroids for eight HST-GSC stars in our central field and 
used these stars to 
derive an approximate plate solution.  Color and magnitude effects on the derived positions have not been accounted 
for.  The coordinates listed in our variable catalogues (Tables I and III) are accurate to within 
approximately 0."5.  A CCD template image and the X-Y positions of the variable stars were submitted 
for electronic availability through the Astronomical Journal.

\section{Light Curve Fitting}

Our UBV data include 56 V band observations of Leo II.  All but one of 
these were deep enough for reasonable photometry of the horizontal branch.  Our 
observations also include thirteen usable B band observations and three usable U band observations, 
which we have excluded from the period fitting due to the very poor phase coverage
of the images.  A file containing the Julian dates and V magnitudes of all observations of our
variables stars is available electronically through the Astronomical Journal.

To identify variables in our data, we used a slightly modified version
of the Welch-Stetson (1993) variability index, adopted
from that calculated in the DAOMASTER code included with ALLFRAME.  We confined our variability 
search to V images, as they were more numerous and of better quality.  Figure 2
shows the index vs. magnitude plot and our selection criteria (Index $\ge$ 3.0).  A limit
of 3.0 was selected as the lowest index where most stars had clear light curves and below
which significant numbers of spurious detections began.
We restricted our attempts to fit light curves to stars that had at least 
twenty observations and had been observed in both runs.

Periods were identified by both Phase Dispersion Minimization (Stellingwerf 1978)
and Stetson's (1996) modified version of the Lafler-Kinman index (1965).  The IRAF
version of PDM proved very useful for identifying true variables and revealing
the nature of those variables.  It was especially adept at narrowing the range within 
which to search for periods, at finding periods for the short-period variables, 
and at producing light curves.
The modified LK method proved more effective with longer period 
stars.  Once a family of reasonable periods was identified, we ran each 
through a $\chi^2$ fitting routine to refine amplitudes and to break the 
degeneracies of equally suitable periods.  This routine used the templates 
of Layden (1998) for RRab stars and a sinusoidal curve for RRc.

In almost all cases, an obvious best fit was obtained.  In some cases, degeneracies 
remained.  These occured in two intervals - large degeneracies of 0.2-0.5 days and small
degeneracies of 0.01-0.02 days.  Large degeneracies could be broken by using the well-established 
differences between RRc, RRab and Cepheid stars (rise time, amplitude and color).  
Small degeneracies were slightly more 
difficult to break.  We settled for the period giving the lowest $\chi^2$ fit.  This period was 
then varied by several $10^{-5}$ days to further improve the fit.

The greatest deficiency of this technique is that a star that is not sampled at or 
near the peak brightness (within 0.1-0.2 in phase) will not be amenable to fitting.  In fact, for 
low amplitude variables, it may 
not even have the contrast in brightness to be flagged as a variable.  We have attempted to increase
the number of detected variables by lowering the variability index threshold to 2.0.  Approximately
10-20 stars in this range are potential RR Lyrae variables based on their PDM spectra.  However, 
the magnitude contrast over the light curve is so slight that solutions prove very degenerate, if
they are obtainable at all.  For some of these stars, different periods have a $\chi^2$ value below
1.0, making them statistically equivalent.  Rather than include a small number of very poor 
or very degenerate variables, we left any variable with an index below 3.0 out of our sample.

Classification of variables was straight-forward.  The three types of variable
stars that commonly occur in dSph galaxies are RR Lyrae type ab, RR Lyrae type c 
and anomalous Cepheids.  All three objects have well-established loci in 
period--amplitude--luminosity--effective temperature--rise time space.

\section{The Variable Star Catalogue}

Our variable star catalogue (Tables I and III) lists two identification numbers.  The
first is a running number for the variables alone, beginning with the 80 candidates from DI93. The second 
is a corresponding number from our master photometry catalogue of Leo II.  These ID numbers are from ALLFRAME
and roughly correlate with magnitude.  Periods, amplitudes and pulsation modes are listed for stars
for which these parameters could be derived.
Magnitudes for most of the variable stars are intensity-weighted mean 
magnitudes.  They were derived by integrating the template light curve in 0.02 phase increments 
(the same phase increments used in the Layden templates).  For variables for which no
variability parameters could be derived, the average observed magnitude is given.

We have compared our catalogue of variable stars to the list of potential variable stars in DI93.  
Of the 80 potential variable stars, we recover 40 as RR Lyrae type ab, 13 as type c and 
six as unknown RR Lyrae--like stars.  One other star is an anomolous Cepheid.  The remaining
fifteen are not variables according to our analysis although seven of these fifteen have variability indices 
between 2.0 and 3.0, which places them just
below our selected variable envelope; these may be true variables of very low amplitude or variables which were 
not in optimal phase during our observations.  Table II
lists the variable candidates in DI93 that did not meet our variability criterion, along with
their corresponding identification number in our master photometry catalogue and our Welch-Stetson variability
index.

\section{The RR Lyrae Variables}

Figure 3 shows the amplitude-period distribution of the positively identified
RR Lyrae variables in Leo II.
Parameters of the variables are given in 
Table I while light curves are shown in Figure 4.

We note the well-established period-amplitude
relationship in the type ab variables.  The c variables are of nearly constant
amplitude although there is some hint of the parabola-like shape predicted
by Bono et al. (1997).

For eight variables with RR Lyrae-like variations
we were unable to obtain satisfactory fits to their photometric data.  These stars could
potentially be Blazhko (1907) variables or double-mode pulsators 
(Cox et al. 1980; Sandage et al. 1981, hereafter SKS; Cox et al. 1983; Nemec 1985a).
Our data are not sufficient for a more rigorous analysis of their light curves.
We have identified an additional three variables that have a well-defined light
curve in one epoch of data and a poorly defined one in the other.  The
likely explanation is that these are exhibiting the Blazhko effect.  
They are listed in Table I with the 
best fit that could be obtained to one epoch of photometric data; these stars are noted
with a classification of ``abb''.

\subsection{Leo II's Stellar Populations}

Sandage (1993a, S93a) demonstrated that the shortest RRab period could be used to
discern the location of the blue fundamental edge of the instability strip, 
which is a function of metallicity.  Using his relation of 

\begin{center}
log($P_{ab}$) = -0.122 [Fe/H] - 0.500
\end{center}

\noindent and the shortest RRab period of 0.50692 days (log(P) = -0.295), we derive a 
metallicity of -1.68 on the Butler-Blanco scale (Butler 1975; Blanco 1992).  This scale
is 0.2 dex richer than the standard Zinn-West (1984) scale, so we correct our metallicity  
to -1.88.\footnote[2]{The Zinn-West scale has been examined by Carretta \& Gratton (1997) and
Rutledge et al. (1997) and has been shown to have non-linearity in comparison to their
metallicity measures.  While metallicity measurements in this paper use the Zinn-West scale, 
the non-linearity should have only a minor effect upon our conclusions.}
This is closer to the low metallicity of DI93 and 
other ground-based studies than the higher MR96 metallicity of [Fe/H]=-1.6.  A missing variable with a 
period of 0.469 days would move the metallicity scale back to the MR96 value.  Such a low value 
does not occur in any of our fitted RRab periods, nor in the degeneracies of our
unfitable variables.

S93a also defined several other relations for metallicity determination, 
which we list in order of declining sensitivity of our data.  The mean 
period of RRab variables is well defined by our data (0.619 $\pm$ 0.006 days).  S93a
defines several relations between average RRab period and metallicity.  We have 
chosen the relationships for cluster variables in which the average period has not 
been corrected for number density across the instability strip.  This relation is 
use parameters closest to the actual measured quantities:

\begin{center}
log$<P_{ab}>$ = -0.092 [Fe/H] - 0.389
\end{center}

\noindent This produces a Zinn-West metallicity of -1.96 $\pm$ 0.04 dex.  The mean period of RRc 
variables (0.363 $\pm 0.008$ days), via the relation:

\begin{center}
log$<P_{c}>$ = -0.119 [Fe/H] - 0.670,
\end{center}

\noindent produces a Zinn-West value of -1.93 $\pm 0.08$ dex.  Finally, the longest RRab period 
defines the red edge of the instability strip and thus the lower bound of metallicty.
Our longest ab period is 0.8081 days.  The relation:

\begin{center}
log($P_{ab}$) = -0.09 [Fe/H] - 0.280
\end{center}

\noindent produces a -2.08 dex (-2.28 on the Zinn-West scale) lower bound for [Fe/H].

The beauty of the Sandage relations is that they are completely independent of 
photometric zero point, reddening or calibration.  Our results are consistent 
with previous metallicity estimates of 
Leo II, which have usually been around [Fe/H]=-1.9 (DH83, S86, DI93, L95).  

The notable exception to these consistent metallicity determinations is MR96.  Their analysis showed that
photometric contamination of the red giant branch by red clump stars could
have shifted DI93's estimated metallicity.
However, MR96 also noted a 0.10 magnitude 
difference between their $V$ magnitudes and those of DI93, which they 
attributed to inaccurate airmass estimates on the part of DI93, possibly as 
a result of the eruption of Mt. Pinatubo.  We are puzzled by this explanation, since DI93 
observed standard stars on the same nights as Leo II.  Unless the distribution of Pinatubo debris were 
extremely inhomogenous, it should affect the extinction terms equally for both 
standard and Leo II stars.  The MR96 explanation would require a conspiracy of higher extinction only during
Leo II observations.
L95 detected a smaller $V$ band difference of 0.02 magnitudes
in the brightness range of interest (both L95 and our own comparison show some non-linearity at the bright
end of the DI93 CCD data).  A zero point correction to the $V$ and $I$ magnitudes of MR96 might 
result in a shift of their color, which would produce a metallicity-reddening 
shift based on the MR96's use of Sarajedini's (1994) calibration technique.  A shift of 0.06 
in color would move the giant branch onto an [Fe/H]=-1.9 metallicity relation.  In the MR96 analysis, 
this is a $6 \sigma$ deviation.  However, Stetson (1998) has shown that WFPC2 data's
best attainable photometric precision is approximately 0.02 magnitudes due to charge-transfer efficiency problems
in the detectors.  
When taken in quadrature
with MR96's (very small) photometric scatter and the uncertainties of the HST zero points 
(Holtzman et al. 1995), we estimate the uncertainties to be more likely near 0.03-0.04 magnitudes.  
This would reduce the shift of MR96 zero points to a $1.5-2 \sigma$ effect.  Given the 
comparisons between the V band magnitudes of MR96, L95, DI93 and our own study, we find
a systematic magnitude shift of the WFPC2 data to be a plausible reconciliation of all the 
metallicity measures of Leo II.

However, the existence of a more metal-rich population could be plausible if it is too young to have a
significant RR Lyrae population.  MR96 showed that the youngest stars in
Leo II are 7 Gyr old, with a typical star having an age of 9 Gyr.  This is
less than what is generally thought to be the minimum age possible for an RR Lyrae
star, 10-12 Gyr (Olszewski et al. 1987, O87).

It is not clear how much weight such an argument should be given.
First, O87 and MR96 use different methods of analysis.  O87 used the $BV$
isochrones of VandenBerg and Bell (1985) and VandenBerg (1985).  MR96
averaged eight different age estimates, ranging from 7 to 11 Gyr (their
Table 5).  Of these, the most comparable age indicator to that used in O87 are the 
Revised Yale Isochrones
(Green et al. 1987) which produced an age of 10 Gyr, just inside the O87
RR Lyrae minimum age envelope.  Second, Leo II is only 1-3 Gyr younger
than the supposed RR Lyrae minimum age, a difference comparable to the
uncertainties (2 Gyr in both cases).
Finally, the lower bound of RR Lyrae ages has only been determined by the
comparison of Lindsay 1 to other LMC clusters and such an important result will not be conclusive 
until a wider variety of clusters lacking RR Lyrae's have been studied.
Until the issues involved with globular cluster ages are fully resolved -
and more cluster ages are determined - the possibility of a young, more metal-rich, 
variable-less population remains open.

We are intrigued by the presence of ab variables with periods longer than what should
be the red edge of an [Fe/H]=-1.9 population.  These imply
a population that is more luminous than the bulk of Leo II variables.
To separate out the more luminous stars in Leo II cleanly, we have applied a period-shift analysis
to our data (Sandage 1981a; Sandage 1981b; Carney et al. 1992).  

Period shift is measured by comparison to a reference variable population.  The usual 
candidate population is that of M3, for which Sandage (1982a, 1982b) derived a measure of 
$\Delta log P = -[0.129 A_{B} + 0.088 + log P]$, where $A_{B}$ is the amplitude in the 
$B$ passband.  However, we applied this formulation to the recent 
measurements of M3 variable periods and amplitudes by Caretta et al. (1999) and 
Ka{\l}u\`zny et al. (1998) 
and found a period shift of 0.015 by the Sandage formulation.
We are unable to explain this discrepancy but have corrected the period
shift zero point to produce a $\Delta log P$ of 0.000 for the new M3 data.  We have also derived
a conversion ratio of $A_B/A_V$=1.21.  The resulting revised period-shift relations are:

\begin{center}
$\Delta log P = -[0.129 A_{B} + 0.112 + log P]$

$\Delta log P = -[0.156 A_{V} + 0.112 + log P]$
\end{center}

\noindent Our $\Delta log P$ values for Leo II from these relations are plotted in Figure 5.

The pulsation equation of Van Aldada \& Baker (1971) is:

\begin{center}
$log P_{o} = -1.772 - 0.68 log (\frac{M}{M_{\sun}}) + 0.84 log (\frac{L}{L_{\sun}}) + 3.48 log (\frac{6500}{T_{eff}})$
\end{center}

\noindent Assuming a uniform mass, one can measure changes in luminosity by measuring the period and effective
temperature of each star.  Effective temperature can not be directly measured, but SKS
showed that amplitude, specifically in the blue passband, can be used to measure
effective temperature.  The most recent revision of this relation is by Catelan (1998):

\begin{center}
$\frac{5040}{T_{eq}} = 0.868 - 0.084 A_{B} + 0.005[Fe/H]$
\end{center}

\noindent Where $T_{eq}$ is the equilibrium temperature.  
Equilibrium temperature is not the same as effective temperature, but is very similar (Carney et al. 1992).  
This relation has only a weak dependence upon metallicity.  By comparing the periods of two stars
with identical amplitudes, we compare stars at identical temperatures.  This enables us to
measure differences in luminosity.

The resultant theoretical $\Delta M_V$ scale is on the right ordinate of Figure 5.  While there is a large 
amount of scatter intrinsic to the period-shift measure (and our conversion of
the relations to the $V$ passband only amplifies this scatter) it is clear that the stars in Leo II 
are dominated by a population 0.08 magnitudes more luminous than the stars of M3 with a higher 
luminosity tail in the longer period variables.

To confirm the increased luminosity for the longer period variables, we have directly compared the apparent 
magnitudes of our
higher period shift population to the bulk population.  
We find that the nine most shifted stars ($\Delta log P < -.08$) in
Figure 5 (excluding the two that clearly deviate from the main magnitude locus in Figure 8) 
average $0.09 \pm .03$ magnitudes brighter than the bulk of the RR Lyrae stars.  This is smaller 
than the 0.2 magnitude difference implied by the period-shift analysis (see Figure 5), which 
indicates a difference in mass for the more extreme Leo II variables.

If we abandon the assumption of fixed mass and use differences in magnitude to constrain 
luminosity differences, we can measure the relative mass of each star by rewriting
the pulsation equation as:

$-0.68 log \frac{M_{avg}}{M} = -0.336 (m-m_{avg}) + \Delta log P - <\Delta log P>$

\noindent where $M_{avg}$ is the average RR Lyrae mass, $M$ is the mass of an 
individual star, $m_{avg}$ and $m$ are the average and individual apparent 
magnitudes, and $<\Delta log P>$ is the average period shift
in Leo II.
Using this formulation, we estimate the second population of stars to be 13\% less massive 
than the average Leo II RRab star.

While longer period variables can occur in the 
absence of metallicity effects (Lee \& Carney, 1999a; Pritzl et al. 1999), this signature is also
seen in Sculptor (Ka{\l}u\`zny et al. 1995, hereafter K95), which has a well-established spread in 
metallicity (Norris \& Bessell 1978; Smith \& Dopita 1983; Da Costa 1984; Da Costa 1988; K95; 
Majewski et al. 1999, hereafter M99; Hurley-Keller et al. 1999).  This may warrant a closer look at the Leo 
II color-magnitude diagram to see if a dual metallicity model may be more appropriate, as in the 
case of Sculptor.  While 
dramatic conclusions should not be drawn from a handful of unusual stars, if 
the increased luminosity is the product of metallicity alone, this second 
Leo II population would have a metallicity of [Fe/H]=-2.3 using the RR 
Lyrae luminosity calibration of Sandage (1993b).

\subsection{Oosterhoff Classification}

It has long been known that Galactic clusters fall into two categories based
upon the properties of their RR Lyrae variables (Oosterhoff 1939; see also a
history of this phenomenon in S93a and Smith 1995).
The ratio of c to ab variables of 0.24, the average RRab period of 
0.62 days, and the minimum RRab period of 0.51 days
are hallmarks of an Oosterhoff classification for Leo II that is intermediate 
between OoI and OoII.  This classification is also seen in the Sculptor 
dSph (K95), Sextans (Mateo et al. 1995), 
several Magellanic clusters (Bono et al. 1994) and possibly the Draco (Nemec 1985b), 
Carina (Saha et al. 1986) and Ursa Minor (Nemec et al. 1988) dSph's.

S93a argued that the Oosterhoff dichotomy 
is the result of a continuous change of RR Lyrae properties with
metallicity.  The gap between Oosterhoff classes would result from Galactic
clusters with metallicites between OoI and OoII having extreme blue horizontal branches
that depopulate the instability strip (Renzini 1983; Castellani 1983).  
Leo II and other Oosterhoff objects, however, have a second parameter effect
that populates the red end of the horizontal branch and the instability strip.
Figure 6 plots the average ab periods and spectroscopic metallicities of the
clusters listed in S93a and the intermediate Oosterhoff 
clusters and dSph's listed above.  The Oosterhoff imtermediate objects fill in the
gap between the Oosterhoff classes and fall close to the relation
of S93a.  This would support the Renzini-Castellani-Sandage
explanation of the Oosterhoff dichotomy.

The one caveat to this conclusion is that the dwarf spheroidal galaxies
show evidence of multiple populations (see a summary in Grebel 1997, Grebel 1998 and Mateo 1998).  
This may be reflected in 
the variables of Leo II as well as those of Sculptor, which has a metallicity
distribution from [Fe/H]=-1.5 to -2.3 (Ka{\l}u\`zny et al. 1995; M99).  In principle, one could
revive the Oosterhoff dichotomy by supposing that Oosterhoff intermediate
objects merely reflect a superposition of OoI and OoII populations.

We find this to be a dubious interpretation.  In the first place, the four intermediate
LMC clusters all clearly have single populations.  Second, in the 
case of Leo II and Sculptor, the consistency between non-variable measures 
of metallicity and the S93a formulation for the RR Lyrae stars indicates that the less metal-poor 
population is the origin of the bulk of the variables.

A third argument against this can be seen in Figure 5.  The period shift is also a 
useful measure of the Oosterhoff effect in individual stars.  Both cluster stars and
field stars exhibit the Oosterhoff dichotomy by avoiding $\Delta log P$ values between 
-0.01 and -0.05 (Suntzeff et al. 1991).  If the Oosterhoff intermediate objects 
are a superposition of OoI and OoII, the mean $\Delta log P$ value may be in the forbidden
zone but individual stars should avoid it.
Clearly, just as the Oosterhoff dichotomy
is exhibited for Galactic cluster and field stars on a star-by-star basis, so do the variables 
of Leo II violate that dichotomy on a star-by-star basis.  Our reformulation
of the $\Delta log P$ measure places 56 of our 106 RR Lyrae 
ab variables into the forbidden region.  Using the classical 
$\Delta log P$ formulation places 39 of our stars in this region.
\footnote[3]{An identical argument is used by Mateo et al. (1995) to 
show that the RRab stars of Sextans are Oosterhoff intermediate.}

This by no means implies that metallicity is the only parameter that 
affects the bulk and individual properties of cluster RR Lyrae stars.  Metallicity
has been suggested as the primary factor because of the likely increase in RR Lyrae
luminosity with declining metallicity, which would produce the Oosterhoff effect 
(SKS).  Any other effect that increases RR Lyrae luminosity, 
including HB evolution, would also affect the properties of the variable stars 
(Lee, Demarque \& Zinn 1990, LDZ).  Evolution is apparently the best explanation for the
deviation of M2 from the primary locus in Figure 6 (Lee \& Carney 1999b).  
However, Figure 6 shows that metallicity is the ``first parameter" of average RRab
period.

\subsection{Distance Modulus}

RR Lyrae variables are a useful tool for measuring distance modulus.  
However, the only studies that could be used to calibrate our data
to true apparent magnitudes (DH83, DI93, L95, MR96) also use the horizontal branch as
a distance indicator.  Adapting these other studies to calibrate our RR Lyrae mean magnitudes for the
same purpose would be circular and pointless.
However, if used in combination with other calibration techniques, our RR 
Lyrae variable photometry might eventually serve to refine Leo II's distance measure.  

More importantly, the metallicity
spread in the RR Lyrae variables of the dSph galaxies could assist with the refinement
of the $M_{V} - [Fe/H]$ relation.  Similar efforts with the globular
cluster $\omega$ Centauri (Dickens 1989) have been inconclusive, possibly because of the complicating
effects of evolution (Gratton et al. 1986; Lee 1991).  It has been argued that there
is no universal $M_{V} - [Fe/H]$ relation that is independent of the effect of evolution
as expressed in horizontal branch morphology (LDZ; 
Clement \& Shelton 1999; Lee \& Carney 1999a; Demarque et al. 1999).  
Untangling the evolution--metallicity--absolute magnitude question, especially in the absence 
of direct spectroscopic measures of RR Lyrae metallicity, is well beyond the scope of this paper.

\section{Anomalous Cepheids}

Swope (1968) identified four anomalous Cepheid variables in Leo II.  We have also
identified four variables with parameters similar to Swope's that are cleanly separated
from the RR Lyrae locus in period--amplitude--magnitude--color--rise time space.  Light curves and 
derived parameters are given in Figure 7 and Table III respectively.  We have listed what we believe
to be the correct cross-identificiations to the anomalous Cepheids of the Swope study.  However, her V1 and V51 
are not clearly distinguishable in the absence of a finding chart.

The apparent magnitudes and periods of all RR Lyrae and Cepheid variables are plotted in Figure 
7 with the fundamental and first overtone Cepheid pulsation lines from Nemec et al. (1994).  
We find three fundamental and one first overtone pulsators among the anomalous Cepheids.

The origin of anomalous cepheids and their implications are still poorly understood 
(c.f. Nemec et al. 1988; Mateo et al. 1995).  They can be indicative of a 5-10 Gyr old
population, which MR96 showed exists in Leo II.  They can also be mass-transfer
binaries, for which a specific ratio of $\sim 1-10$ for blue-stragglers to anomalous Cepheids is
predicted (Renzini, Mengel \& Sweigart 1977).  Our photometry of 
Leo II, at present, does not allow a robust estimate of the blue straggler content due to the
rapidly decling photometric accuracy as the data approach the main-sequence
turnoff.  
However, if the suggested ratio were to hold up, the deep photometry of MR96 would be expected
to show 0.5-5 blue stragglers (normalizing to the ratio of horizontal branch stars in our study
and MR96).  Even a cursory glance at 
their Figure 4 reveals a much higher number of blue stragglers.  However, it is unclear
as to how centrally concentrated these objects are (the MR96 pointing is close to the center
of Leo II).  Only deep wide-field photometry of
the entire dSph will give an accurate statistical handle on this question.

\section{Conclusions}

\subsection{Leo II and the Oosterhoff Continuum}

The RR Lyrae variables in Leo II place its dominant metallicity at
[Fe/H]=-1.9.  A handful of large $\Delta log P$ stars appears to imply a lower metallicity
population.  This is not conclusive, but warrants a more comprehensive 
look at the color-magnitude diagram of Leo II, which we will perform in a future 
contribution.  At present, we can make no contribution
regarding Leo II's distance modulus due to the non-photometric conditions 
during our observations.

Leo II is Oosterhoff intermediate class, like several other objects, including
a number of dSph galaxies.  Its 
violation of the Oosterhoff gap is exhibited in both the bulk and individual
properties of its RR Lyrae stars.  The Oosterhoff intermediate objects
fill in the Oosterhoff gap, and therefore support the interpretation that metallicity is the 
dominant parameter in determining average RRab period.

\subsection{The Fornax-Leo-Sculptor Stream Revisited}

Lynden-Bell (1982) was the first to note that the dSph galaxies appear
to lie in two great streams in the sky.  One of these streams appears to 
include the Fornax, Leo I, Leo II and Sculptor dSph galaxies.  Majewski (1992) noted 
that Phoenix, Sextans and several second parameter objects also lie along
this plane.  Palma et al. (2000) found that the outermost ($R_{GC} > 25$ kpc) second
parameter globular clusters have the highest probability of being aligned with
one of the two streams.
While such alignments could be coincidental, they could 
also result from a common formation mechanism, in which case one might expect
some similiarities in the stellar populations of brethren objects.

Comparisons between the populations of Leo II and Sculptor are possible
from the available data.  Both show a strong low metallicity ([Fe/H]=-1.7
in Sculptor), second parameter, Oosterhoff intermediate population.  
Both also show evidence of a less populous, even more metal-poor
population.  In Sculptor, this population does not have a 
second parameter effect and is spatially dispersed (M99).  The 
presence of a spatial gradient in the horizontal morphology of Leo II (Da Costa et al. 1996)
may indicate a similar abundance-HB morphology pattern to that observed
in Sculptor. 

Fornax also has multiple populations.  
While it is dominated by a population at [Fe/H]=-1.5, it shows 
evidence of populations from -0.7 to -2.2 in 
both its field star and globular cluster population (c.f. Buonanno et al. 1985; 
Beauchamp et al. 1995).  The most metal-poor population, like that in Sculptor, 
is much more extended than the metal-rich populations (Grebel \& Stetson 1998).
One could envision a scenario in which Fornax, Leo and Sculptor all 
originated in a common [Fe/H]$\sim$-2.3, first parameter HB progenitor, then, after
disociation, each object followed it own star formation history.  Alternatively, perhaps
there is something fundamental about the ``threshold" metallicity [Fe/H]=-2.3 in small stellar
systems.  We note that the lowest metallicities in the Milky Way globular cluster system are also
around [Fe/H]=-2.3 (Harris 1996).

Any firm conclusions about correlations between Leo II, Sculptor 
and Fornax must await the measurement of their absolute proper motions.  
This will be the single most powerful discriminant for or against a
common history.  However, the similarities between the most 
metal-poor populations in each of Leo II, Sculptor and Fornax are intriguing.

\acknowledgements
The authors would like to thank Marcio Catelan for reading this article before 
submission and providing many useful comments.  We would also like to thank the anonymous
referee for useful remarks.  MHS was provided thesis travel support by NOAO.  
MHS and SRM were supported by awards from the David and Lucile Packard Foundation and the 
Research Corporation, as well as National Science Foundation CAREER Award grant, AST-9702521.

\figcaption[Siegel.RRfig1.gif]{Comparison of stars from DI93 to the corrected $B$ and $V$ magnitudes
of this study.  While the comparisons have RMS values of 0.01 magnitudes, both show a degree
of non-linearity in comparison.}

\figcaption[Siegel.RRfig2.gif]{$V$ magnitude plotted against the modified Welch-Stetson index.  All
objects above the dashed line were marked as potential variables.  While the bright end locus
moves above this line, almost all of these objects had increase indices due to saturation effects.}

\figcaption[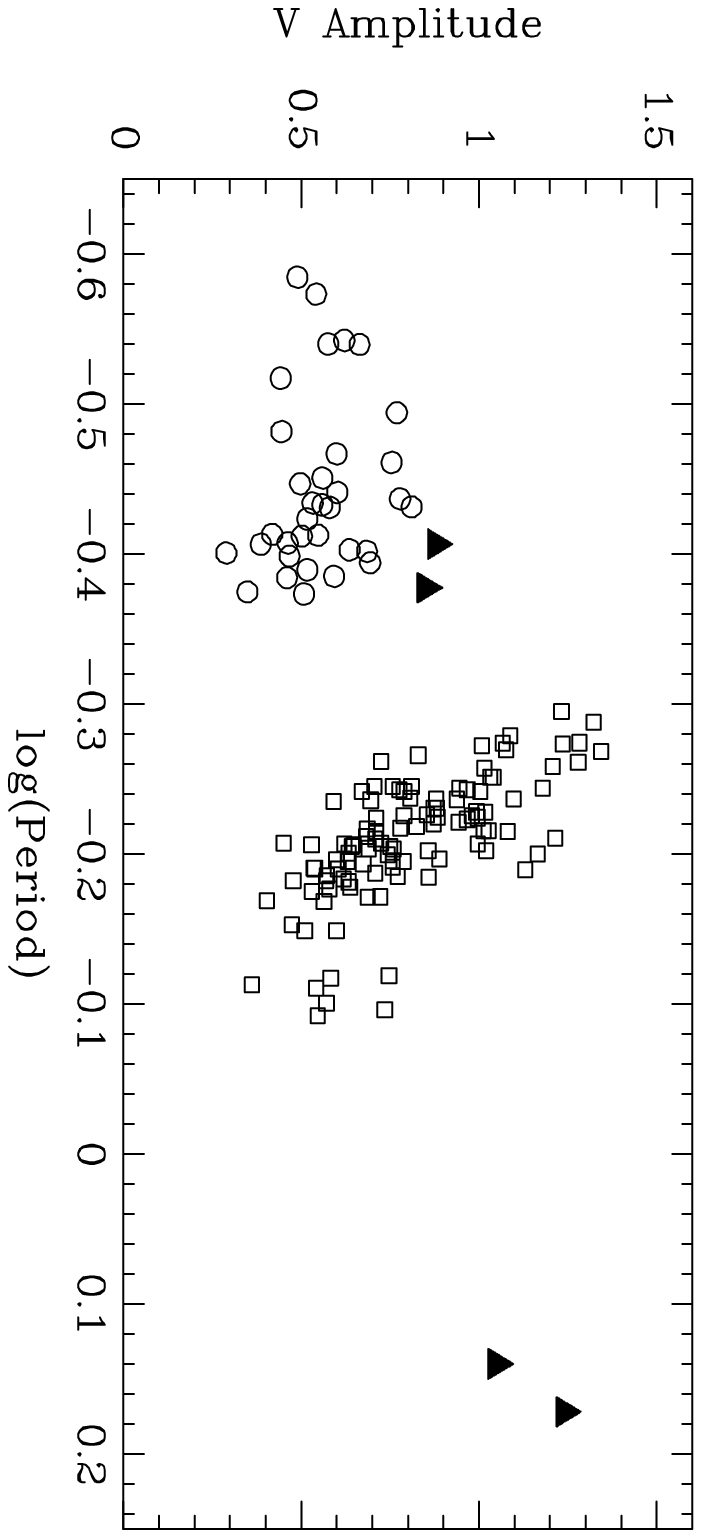]{Period-amplitude distribution of the variables in Leo II.  
{\it Boxes} are RRab variables, {\it circles} are RRc, {\it filled triangles} are anomalous Cepheids.  
Periods are in days.}

\figcaption[Siegel.RRpage1.gif]{RR Lyrae light curves in order of increasing period.}

\figcaption[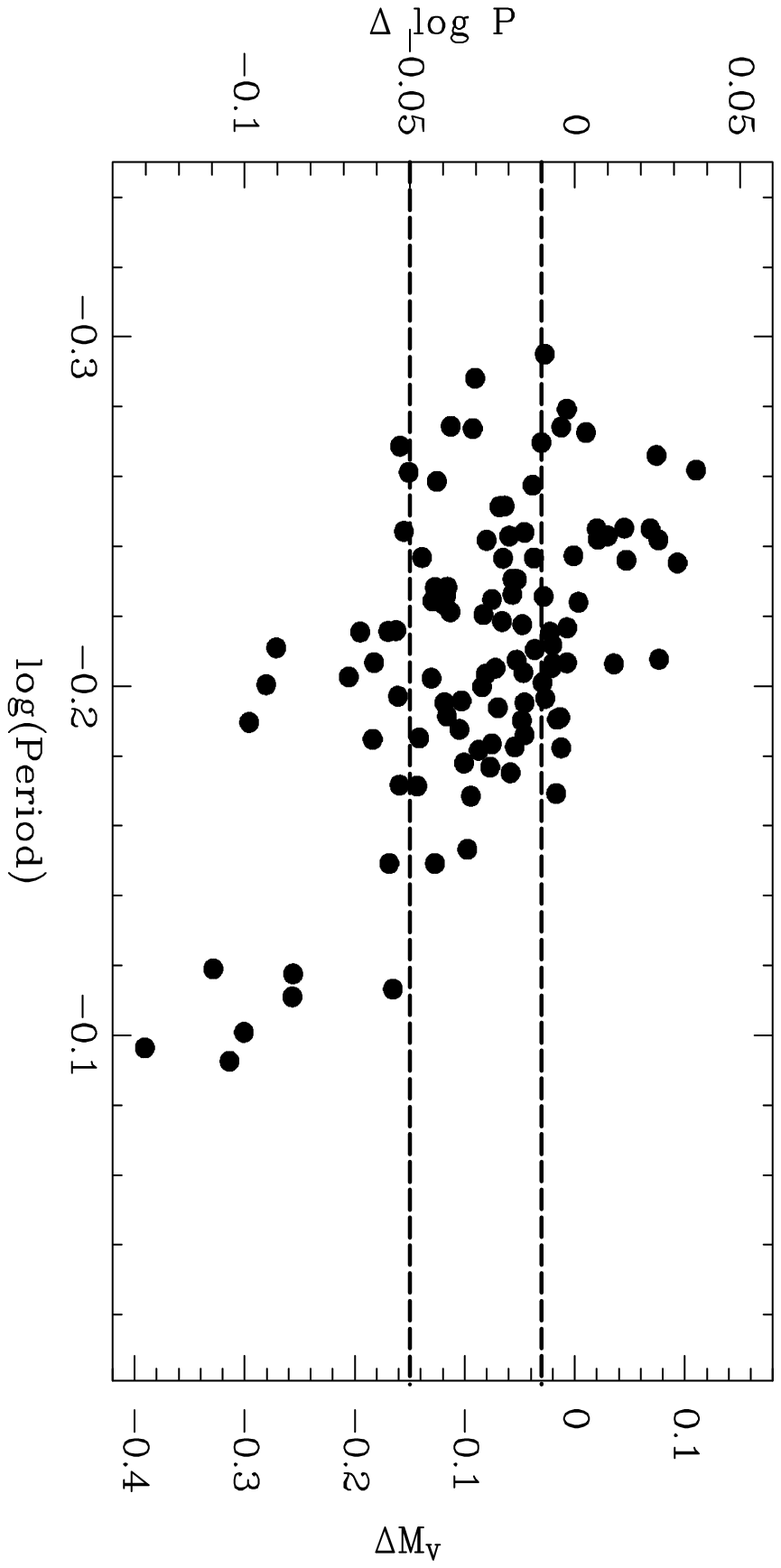]{The period shift effect in the variables of Leo II.  
$\Delta log P$ is the period shift.  The right
ordinate shows the corresponding magnitude shift, under the assumption of constant
mass.  The dashed lines mark the 
zone avoided by Galactic and cluster stars.}

\figcaption[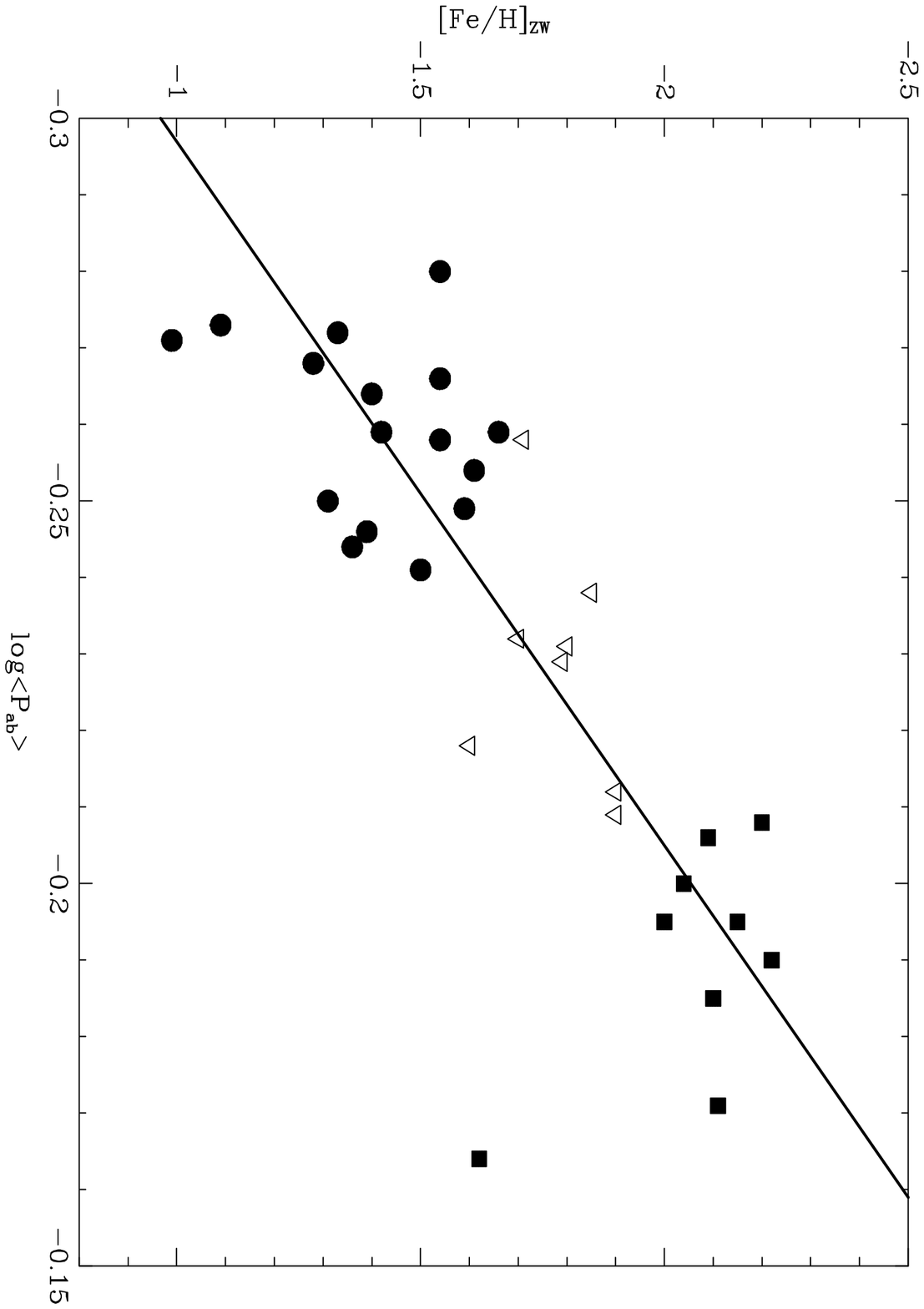]{Average RRab period against spectroscopic metallicity for globular clusters and dwarf 
spheroidal galaxies with 10 or more measured RRab variables.  {\it Circles} are OoI clusters, {\it open 
triangles} Oo intermediate and {\it squares} OoII.  For this plot, we have adopted Draco as Oo 
intermediate, Carina and Ursa Minor as OoII.  The solid line is the observational fit from Sandage 
(1993a).  The significantly deviant point is M2, based on the study of Lee \& Carney (1999b).}

\figcaption[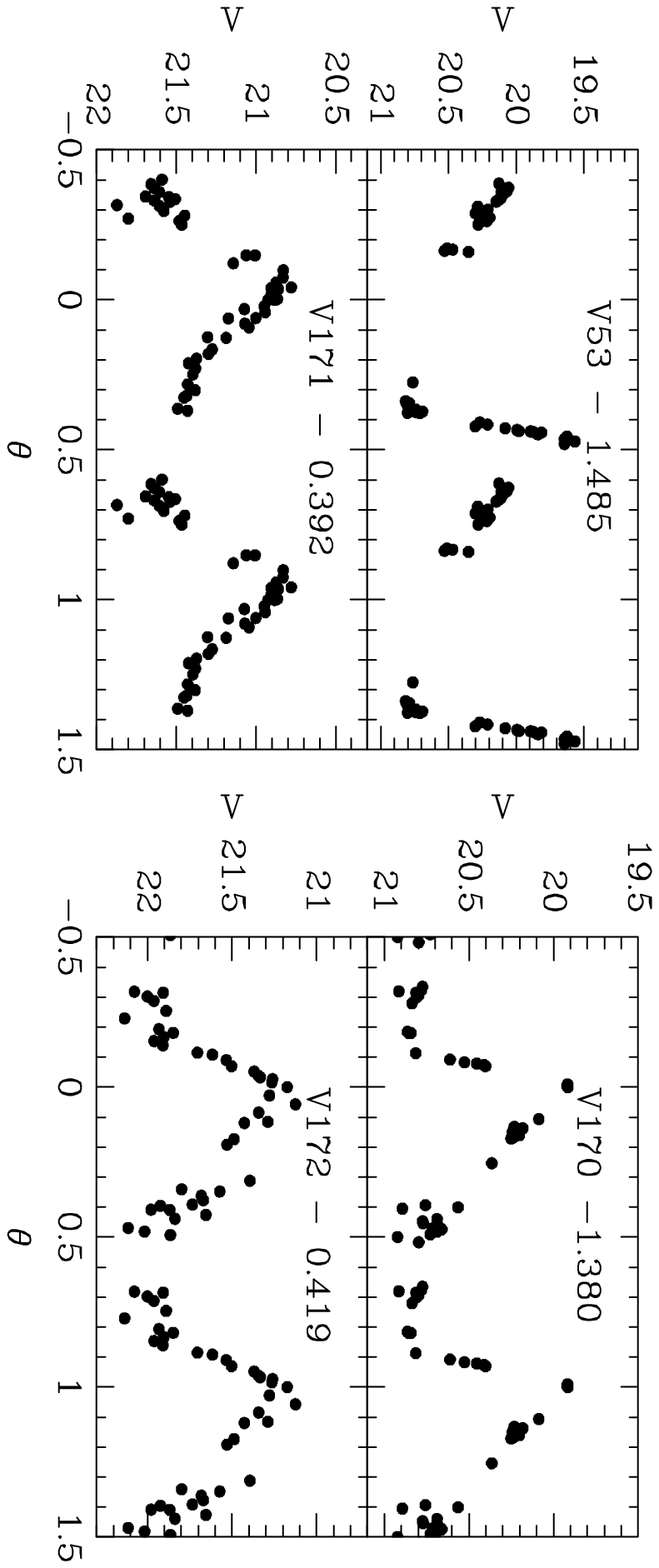]{Light curves of the anomalous Cepheid variables in Leo II.}

\figcaption[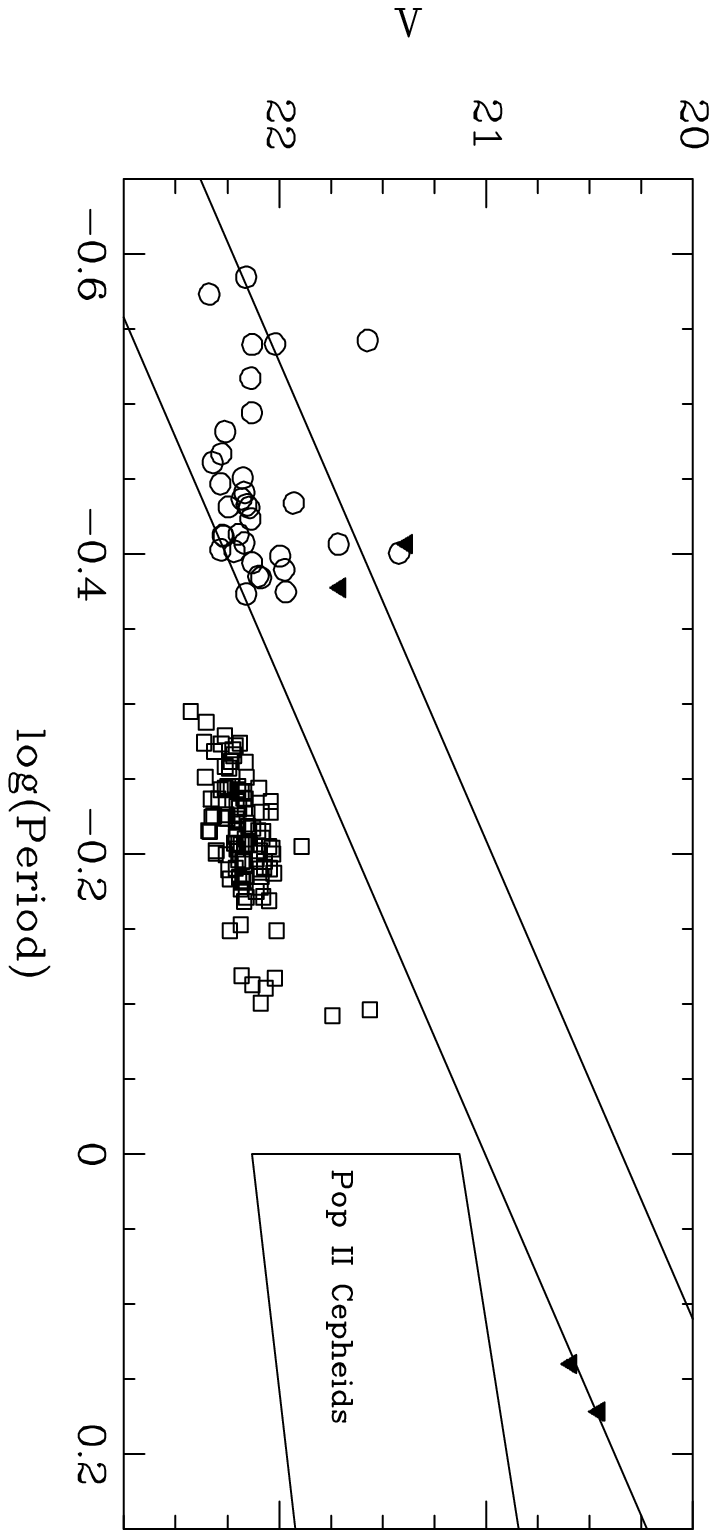]{Apparent magnitudes of all variables stars within Leo II as
a function of period.  
RRab variables are marked with {\it squares}, RRc with {\it circles} and anomalous Cepheids with 
{\it filled triangles}.  
The solid lines are the fundamental and first overtone modes produced by Nemec et al. (1994).}

\begin{center}
TABLE I. RR Lyrae Variables\\
\begin{tabular}{c|c|c|c|c|c|c|c|c} \tableline\tableline
Var ID & Star ID &  RA         &  DEC      & Period  & $A_{V}$ & $m_{V}$  & Bailey & $\Delta log P$\\
       &         &  J2000.0    & J2000.0   & (Days)  &         &          & Type   &              \\
\tableline
V1 & 2350   &  11:13:23.36  &  22:12:56.5  &  0.67352  &  0.723  &  22.08 	&  ab  &  -0.053\\
V2 & 2755   &  11:13:32.25  &  22:12:13.1  &  0.62137  &  0.622  &  22.17 	&  ab  &  -0.002\\
V5 & 4035   &  11:13:39.17  &  22:11:21.7  &  0.57152  &  0.967  &  22.20 	&  ab  &  -0.020\\
V6 & 2597   &  11:13:42.30  &  22:10:58.6  &  0.64544  &  0.604  &  22.09 	&  ab  &  -0.016\\
V8 & 2768   &  11:13:32.68  &  22:11:29.5  &  0.63715  &  0.757  &  22.17 	&  ab  &  -0.034\\
V9 & 2153   &  11:13:27.92  &  22:11:43.3  &  0.62357  &  0.751  &  21.89 	&  ab  &  -0.024\\
V10 & 3149  &  11:13:20.75  &  22:12:02.0  &  0.32033  &  0.769  &  22.13 	&  c   &  \\
V11 & 3373  &  11:13:29.65  &  22:11:23.4  &  0.65546  &  0.619  &  22.24 	&  ab  &  -0.025\\
V14 & 3735  &  11:13:25.80  &  22:11:38.0  &  0.64419  &  0.535  &  22.21 	&  ab  &  -0.004\\
V15 & 2059  &  11:13:31.82  &  22:11:10.6  &  0.62126  &  0.997  &  22.16 	&  ab  &  -0.061\\
V16 & 1892  &  11:13:35.79  &  22:10:53.0  &  0.39198  &  0.386  &  21.71 	&  c   &  \\
V17 & 3262  &  11:13:41.16  &  22:10:18.9  &  0.56056  &  1.040  &  22.36 	&  ab  &  -0.023\\
V18 & 2379  &  11:13:19.63  &  22:11:46.2  &  0.79279  &  0.571  &  22.09 	&  ab  &  -0.100\\
V19 & 3316  &  11:13:34.28  &  22:10:37.0  &  0.40336  &  0.694  &  22.13 	&  c   &  \\
V21 & 3217  &  11:13:38.10  &  22:10:11.8  &  0.65821  &  0.633  &  22.18 	&  ab  &  -0.029\\
V22 & 1965  &  11:13:16.90  &  22:11:36.6  &  0.59125  &  1.017  &  22.08 	&  ab  &  -0.042\\
V23 & 3144  &  11:13:41.58  &  22:09:40.3  &  0.53865  &  1.344  &  22.31 	&  ab  &  -0.053\\
V24 & 4297  &  11:13:24.46  &  22:10:53.0  &  0.39538  &  0.636  &  22.28 	&  c   &  \\
V25 & 3068  &  11:13:20.36  &  22:11:10.3  &  0.38665  &  0.547  &  22.27 	&  c   &  \\
V28 & 3193  &  11:13:32.64  &  22:09:33.2  &           &         &  22.10 	&  unk  &  \\
V29 & 2451  &  11:13:32.27  &  22:09:27.8  &  0.76034  &  0.747  &  22.18 	&  ab  &  -0.110\\
V30 & 3230  &  11:13:30.71  &  22:09:17.3  &  0.57145  &  0.776  &  22.28 	&  ab  &  0.010\\
V31 & 2676  &  11:13:35.88  &  22:08:47.6  &  0.36563  &  0.778  &  22.18 	&  c   &  \\
V32 & 3992  &  11:13:19.63  &  22:09:55.9  &  0.38714  &  0.502  &  22.27 	&  c   &  \\
V33 & 3273  &  11:13:28.22  &  22:09:17.8  &  0.63122  &  0.743  &  22.26 	&  ab  &  -0.028\\
V34 & 2472  &  11:13:18.82  &  22:09:57.0  &  0.62329  &  0.643  &  22.05 	&  ab  &  -0.007\\
\tableline
\end{tabular}
\end{center}
\begin{center}
TABLE I. Continued. RR Lyrae Variables\\
\begin{tabular}{c|c|c|c|c|c|c|c|c} \tableline\tableline
Var ID & Star ID &  RA         &  DEC      & Period  & $A_{V}$ & $m_{V}$  & Bailey & $\Delta log P$\\
       &         &  J2000.0    & J2000.0   & (Days)  &         &          & Type   &              \\
\tableline
V36 & 2770  &  11:13:26.43  &  22:09:14.0  &  0.61411  &  0.683  &  22.21 	&  ab  &  -0.007\\
V37 & 4140  &  11:13:26.46  &  22:09:11.3  &  0.35721  &  0.497  &  22.28 	&  c   &  \\
V38 & 3306  &  11:13:27.06  &  22:09:05.0  &  0.26012  &  0.489  &  22.16 	&  c   &  \\
V39 & 3597  &  11:13:33.00  &  22:08:35.9  &  0.61989  &  0.450  &  22.20 	&  ab  &  0.025\\
V40 & 2863  &  11:13:18.29  &  22:09:38.4  &  0.70926  &  0.510  &  22.01 	&  ab  &  -0.042\\
V41 & 2598  &  11:13:29.06  &  22:08:48.6  &  0.59113  &  0.993  &  22.04 	&  ab  &  -0.039\\
V43 & 3018  &  11:13:32.61  &  22:08:17.2  &  0.56856  &  0.758  &  22.23 	&  ab  &  0.015\\
V44 & 3092  &  11:13:34.95  &  22:07:59.2  &  0.36896  &  0.559  &  22.16 	&  c   &  \\
V46 & 4203  &  11:13:30.18  &  22:08:17.0  &  0.60083  &  0.942  &  22.21 	&  ab  &  -0.038\\
V48 & 3159  &  11:13:24.68  &  22:08:37.9  &  0.70922  &  0.599  &  22.24 	&  ab  &  -0.056\\
V49 & 2322  &  11:13:19.43  &  22:08:59.4  &  0.76280  &  0.583  &  22.02 	&  ab  &  -0.085\\
V50 & 2736  &  11:13:16.78  &  22:09:08.6  &           &         &  22.00 	&  unk  &  \\
V52 & 2409  &  11:13:23.49  &  22:08:35.1  &  0.55269  &  1.015  &  22.24 	&  ab  &  -0.013\\
V54 & 4201  &  11:13:32.64  &  22:07:48.3  &  0.59702  &  0.711  &  22.26 	&  ab  &  0.001\\
V55 & 2695  &  11:13:27.61  &  22:08:06.6  &  0.62718  &  1.020  &  22.30 	&  ab  &  -0.069\\
V56 & 4058  &  11:13:28.52  &  22:08:01.5  &  0.34118  &  0.599  &  22.28 	&  c   &  \\
V57 & 2696  &  11:13:17.33  &  22:08:43.8  &  0.42179  &  0.348  &  21.97 	&  c   &  \\
V60 & 4354  &  11:13:27.22  &  22:07:51.8  &  0.60862  &  1.027  &  22.34 	&  ab  &  -0.057\\
V61 & 2712  &  11:13:29.18  &  22:07:34.6  &  0.56883  &  0.810  &  22.20 	&  ab  &  0.007\\
V62 & 2690  &  11:13:12.35  &  22:08:44.4  &  0.36202  &  0.601  &  22.17 	&  c   &  \\
V64 & 4225  &  11:13:21.44  &  22:07:54.2  &           &         &  22.29 	&  unk &  \\
V65 & 3354  &  11:13:30.42  &  22:07:12.6  &  0.60716  &  0.686  &  22.21 	&  ab  &  -0.002\\
V67 & 2459  &  11:13:17.38  &  22:07:54.3  &  0.56032  &  1.031  &  22.16 	&  ab  &  -0.021\\
V69 & 3207  &  11:13:35.74  &  22:06:23.9  &           &         &  22.12 	&  unk  &  \\
V70 & 3835  &  11:13:27.33  &  22:06:56.8  &  0.62026  &  0.725  &  22.22 	&  ab  &  -0.018\\
V73 & 3236  &  11:13:13.22  &  22:07:48.7  &           &         &  22.11 	&  unk  &  \\
\tableline
\end{tabular}
\end{center}
\begin{center}
TABLE I. Continued. RR Lyrae Variables\\
\begin{tabular}{c|c|c|c|c|c|c|c|c} \tableline\tableline
Var ID & Star ID &  RA         &  DEC      & Period  & $A_{V}$ & $m_{V}$  & Bailey & $\Delta log P$\\
       &         &  J2000.0    & J2000.0   & (Days)  &         &          & Type   &              \\
\tableline
V74 & 4029  &  11:13:27.42  &  22:06:39.2  &  0.59392  &  0.854  &  22.26 	&  ab  &  -0.019\\
V75 & 3277  &  11:13:31.50  &  22:06:16.6  &  0.57313  &  1.003  &  22.17 	&  ab  &  -0.027\\
V76 & 2669  &  11:13:10.62  &  22:07:47.1  &           &         &  22.06 	&  unk  &  \\
V77 & 3611  &  11:13:10.70  &  22:07:43.8  &  0.65676  &  0.569  &  22.18 	&  ab  &  -0.018\\
V78 & 4088  &  11:13:20.34  &  22:07:00.0  &  0.56884  &  0.706  &  22.25 	&  ab  &  0.023\\
V79 & 3548  &  11:13:29.42  &  22:06:08.9  &  0.63790  &  0.787  &  22.07 	&  ab  &  -0.040\\
V80 & 3875  &  11:13:34.82  &  22:05:33.7  &  0.57016  &  0.944  &  22.10 	&  ab  &  -0.015\\
V81 & 1517  &  11:13:07.99  &  22:08:40.7  &  0.80097  &  0.735  &  21.56 	&  ab  &  -0.130\\  
V82 & 1675  &  11:13:27.46  &  22:07:27.7  &  0.39731  &  0.289  &  21.42 	&  c   &  \\  
V83 & 1770  &  11:13:36.13  &  22:15:51.1  &           &         &  21.57 	&  unk  &  \\  
V84 & 1784  &  11:13:43.00  &  22:07:26.4  &  0.60829  &  1.013  &  22.10 	&  ab  &  -0.054\\  
V85 & 1900  &  11:13:11.76  &  22:09:13.9  &  0.60888  &  1.080  &  22.08 	&  ab  &  -0.065\\  
V86 & 1922  &  11:13:43.72  &  22:10:48.4  &  0.28667  &  0.621  &  21.57 	&  c   &  \\  
V87 & 1982  &  11:13:33.26  &  22:10:23.6  &  0.53243  &  1.235  &  22.28 	&  abb  &  -0.031\\  
V88 & 1996  &  11:13:52.92  &  22:08:53.0  &  0.80806  &  0.546  &  21.74 	&  ab  &  -0.105\\  
V89 & 2025  &  11:13:48.63  &  22:10:13.3  &  0.56978  &  1.180  &  22.26 	&  ab  &  -0.052\\  
V90 & 2188  &  11:13:23.76  &  22:15:28.2  &  0.55128  &  1.208  &  22.26 	&  ab  &  -0.042\\  
V91 & 2230  &  11:13:15.20  &  22:04:26.6  &  0.36791  &  0.532  &  21.93 	&  c   &  \\  
V92 & 2269  &  11:13:16.10  &  22:03:55.2  &  0.39916  &  0.467  &  21.99 	&  c   &  \\  
V93 & 2290  &  11:13:44.23  &  22:05:27.8  &  0.28849  &  0.664  &  22.13 	&  c   &  \\  
V94 & 2300  &  11:13:13.55  &  22:05:53.4  &  0.60191  &  0.873  &  22.16 	&  ab  &  -0.028\\  
V95 & 2312  &  11:13:39.20  &  22:07:15.4  &  0.64360  &  0.757  &  22.07 	&  ab  &  -0.039\\  
V96 & 2334  &  11:13:19.34  &  22:07:55.6  &  0.40767  &  0.517  &  21.98 	&  c   &  \\  
V97 & 2382  &  11:13:40.25  &  22:07:43.3  &  0.37708  &  0.517  &  22.14 	&  c   &  \\  
V98 & 2394  &  11:13:39.79  &  22:07:45.7  &  0.58175  &  0.591  &  22.04 	&  ab  &  0.031\\  
V99 & 2448  &  11:13:37.13  &  22:13:44.0  &  0.54193  &  0.829  &  22.22 	&  ab  &  0.025\\  
\tableline
\end{tabular}
\end{center}
\begin{center}
TABLE I. Continued. RR Lyrae Variables\\
\begin{tabular}{c|c|c|c|c|c|c|c|c} \tableline\tableline
Var ID & Star ID &  RA         &  DEC      & Period  & $A_{V}$ & $m_{V}$  & Bailey & $\Delta log P$\\
       &         &  J2000.0    & J2000.0   & (Days)  &         &          & Type   &              \\
\tableline
V100 & 2467  &  11:13:18.04  &  22:07:58.8  &  0.51503  &  1.323  &  22.35 	&  ab  &  -0.030\\  
V101 & 2497  &  11:13:49.69  &  22:08:35.3  &  0.60588  &  0.779  &  22.13 	&  ab  &  -0.016\\  
V102 & 2561  &  11:13:15.43  &  22:10:09.7  &  0.52567  &  1.088  &  22.26 	&  ab  &  -0.002\\  
V103 & 2593  &  11:13:21.09  &  22:11:14.8  &  0.59649  &  0.997  &  22.25 	&  ab  &  -0.043\\  
V104 & 2607  &  11:13:40.42  &  22:08:34.6  &  0.77449  &  0.542  &  22.06 	&  ab  &  -0.086\\  
V105 & 2621  &  11:13:33.61  &  22:04:11.7  &  0.53191  &  1.066  &  22.19 	&  ab  &  -0.004\\  
V106 & 2633  &  11:13:21.18  &  22:10:59.2  &  0.67730  &  0.403  &  22.05 	&  ab  &  -0.006\\  
V107 & 2688  &  11:13:07.32  &  22:07:02.3  &  0.63038  &  1.166  &  22.03 	&  ab  &  -0.093\\  
V108 & 2716  &  11:13:48.58  &  22:15:11.1  &  0.62602  &  0.759  &  22.21 	&  ab  &  0.027\\  
V109 & 2767  &  11:13:22.05  &  22:11:29.5  &  0.58815  &  0.872  &  22.19 	&  ab  &  -0.018\\  
V110 & 2801  &  11:13:44.87  &  22:03:39.7  &  0.64623  &  1.130  &  22.25 	&  ab  &  -0.099\\  
V111 & 2820  &  11:13:28.25  &  22:07:08.2  &  0.28828  &  0.576  &  22.02 	&  c   &  \\  
V112 & 2830  &  11:13:14.64  &  22:09:38.2  &  0.34577  &  0.754  &  22.32 	&  c   &  \\  
V113 & 2871  &  11:13:33.23  &  22:12:45.3  &  0.67845  &  0.564  &  22.17 	&  ab  &  -0.032\\  
V114 & 2887  &  11:13:49.59  &  22:04:44.7  &  0.64921  &  0.709  &  22.03 	&  ab  &  -0.035\\  
V115 & 2896  &  11:13:29.42  &  22:08:23.6  &  0.66809  &  0.530  &  22.11 	&  ab  &  -0.020\\  
V116 & 2899  &  11:13:30.10  &  22:01:49.9  &  0.61526  &  1.214  &  22.08 	&  ab  &  -0.090\\  
V117 & 2906  &  11:13:35.27  &  22:07:34.5  &  0.64489  &  0.538  &  22.05 	&  ab  &  -0.005\\  
V118 & 2943  &  11:13:40.84  &  22:04:01.1  &  0.35405  &  0.559  &  22.18 	&  c   &  \\  
V119 & 2962  &  11:13:24.90  &  22:11:02.5  &  0.77051  &  0.361  &  22.13 	&  ab  &  -0.055\\  
V120 & 2990  &  11:13:15.84  &  22:12:59.7  &  0.65160  &  0.572  &  22.16 	&  ab  &  -0.015\\  
V121 & 3006  &  11:13:36.99  &  22:08:59.5  &  0.62537  &  0.689  &  22.03 	&  ab  &  -0.016\\  
V122 & 3031  &  11:13:28.43  &  22:04:49.6  &  0.53380  &  1.008  &  22.21 	&  ab  &  0.003\\  
V123 & 3055  &  11:13:46.00  &  22:07:26.9  &  0.63797  &  0.631  &  22.18 	&  abb  &  -0.015\\  
V124 & 3072  &  11:13:42.68  &  22:06:25.4  &  0.41264  &  0.460  &  22.09 	&  c   &  \\  
V125 & 3086  &  11:13:39.67  &  22:11:59.1  &  0.65294  &  0.771  &  22.09 	&  ab  &  -0.047\\  
\tableline
\end{tabular}
\end{center}
\begin{center}
TABLE I. Continued. RR Lyrae Variables\\
\begin{tabular}{c|c|c|c|c|c|c|c|c} \tableline\tableline
Var ID & Star ID &  RA         &  DEC      & Period  & $A_{V}$ & $m_{V}$  & Bailey & $\Delta log P$\\
       &         &  J2000.0    & J2000.0   & (Days)  &         &          & Type   &              \\
\tableline
V126 & 3121  &  11:13:20.53  &  22:07:13.3  &  0.58798  &  0.882  &  22.17 	&  ab  &  -0.019\\  
V127 & 3122  &  11:13:03.71  &  22:05:51.8  &  0.37047  &  0.580  &  22.15 	&  c   &  \\  
V128 & 3166  &  11:13:30.67  &  22:09:12.5  &  0.62768  &  0.857  &  22.10 	&  ab  &  -0.043\\  
V129 & 3180  &  11:13:30.03  &  22:05:22.6  &  0.54800  &  1.279  &  22.16 	&  ab  &  -0.050\\  
V130 & 3218  &  11:13:19.76  &  22:06:16.2  &           &         &  22.12 	&  unk  &  \\  
V131 & 3231  &  11:13:06.56  &  22:02:41.1  &  0.63519  &  0.889  &  22.11 	&  ab  &  -0.054\\  
V132 & 3240  &  11:13:36.35  &  22:09:04.3  &  0.32976  &  0.444  &  22.26 	&  c   &  \\  
V133 & 3257  &  11:13:40.16  &  22:06:16.4  &  0.66382  &  0.638  &  22.09	&  ab  &  -0.034\\  
V134 & 3351  &  11:13:11.72  &  22:08:54.0  &  0.41174  &  0.592  &  22.10 	&  c   &  \\  
V135 & 3355  &  11:13:48.37  &  22:06:15.5  &  0.59801  &  0.967  &  22.21 	&  ab  &  -0.040\\  
V136 & 3363  &  11:13:34.75  &  22:07:44.2  &  0.58072  &  0.695  &  22.21 	&  ab  &  0.016\\  
V137 & 3419  &  11:13:32.00  &  22:10:50.2  &  0.54721  &  0.725  &  22.23 	&  ab  &  0.037\\  
V138 & 3470  &  11:13:46.51  &  22:12:51.0  &  0.57274  &  0.789  &  22.18 	&  ab  &  0.007\\  
V139 & 3603  &  11:13:39.59  &  22:12:06.3  &  0.37020  &  0.810  &  22.25 	&  c   &  \\  
V140 & 3612  &  11:13:38.01  &  22:06:10.5  &  0.57298  &  0.670  &  22.2 	&  abb  &  0.025\\  
V141 & 3648  &  11:13:12.49  &  22:04:50.4  &  0.53736  &  1.076  &  22.22 	&  ab  &  -0.010\\  
V142 & 3655  &  11:13:40.13  &  22:10:37.0  &  0.57900  &  0.806  &  22.17 	&  ab  &  -0.000\\  
V143 & 3668  &  11:13:42.93  &  22:07:37.7  &  0.67390  &  0.687  &  22.16 	&  ab  &  -0.048\\  
V144 & 3728  &  11:13:43.62  &  22:06:53.2  &  0.60467  &  0.824  &  22.15 	&  ab  &  -0.022\\  
V145 & 3753  &  11:13:23.40  &  22:06:11.7  &  0.30380  &  0.442  &  22.14 	&  c   &  \\  
V146 & 3756  &  11:13:10.54  &  22:04:10.3  &  0.62178  &  0.529  &  22.14 	&  ab  &  0.012\\  
V147 & 3767  &  11:13:20.15  &  22:05:44.6  &  0.39119  &  0.462  &  22.17 	&  c   &  \\  
V148 & 3793  &  11:13:22.04  &  22:10:54.7  &  0.59595  &  0.884  &  22.33 	&  ab  &  -0.025\\  
V149 & 3819  &  11:13:17.59  &  22:03:32.8  &  0.59440  &  0.980  &  22.20	&  ab  &  -0.039\\  
V150 & 3827  &  11:13:44.15  &  22:14:33.2  &  0.63607  &  0.599  &  22.20 	&  ab  &  -0.009\\  
V151 & 3833  &  11:13:46.82  &  22:09:44.5  &  0.64002  &  0.674  &  22.17 	&  ab  &  -0.023\\  
\tableline
\end{tabular}
\end{center}
\begin{center}
TABLE I. Continued. RR Lyrae Variables\\
\begin{tabular}{c|c|c|c|c|c|c|c|c} \tableline\tableline
Var ID & Star ID &  RA         &  DEC      & Period  & $A_{V}$ & $m_{V}$  & Bailey & $\Delta log P$\\
       &         &  J2000.0    & J2000.0   & (Days)  &         &          & Type   &              \\
\tableline
V152 & 3834  &  11:13:10.49  &  22:07:59.2  &  0.39628  &  0.685  &  22.22 	&  c   &  \\  
V153 & 3855  &  11:13:13.11  &  22:06:31.2  &  0.53161  &  1.282  &  22.36 	&  ab  &  -0.038\\  
V154 & 3858  &  11:13:27.63  &  22:07:37.0  &  0.57966  &  1.097  &  22.16 	&  ab  &  -0.046\\  
V155 & 3899  &  11:13:37.02  &  22:10:09.5  &  0.57998  &  0.938  &  22.29 	&  ab  &  -0.022\\  
V156 & 3922  &  11:13:21.68  &  22:12:57.0  &  0.66558  &  0.580  &  22.19 	&  ab  &  -0.026\\  
V157 & 3943  &  11:13:32.38  &  22:07:20.5  &  0.70267  &  0.473  &  22.19 	&  ab  &  -0.033\\  
V158 & 3949  &  11:13:37.75  &  22:09:35.5  &  0.38603  &  0.418  &  22.20 	&  c   &  \\  
V159 & 3970  &  11:13:28.80  &  22:12:35.5  &  0.62165  &  0.648  &  22.18 	&  ab  &  -0.007\\  
V160 & 3989  &  11:13:18.88  &  22:05:50.7  &  0.65722  &  0.477  &  22.16 	&  ab  &  -0.004\\  
V161 & 4055  &  11:13:18.78  &  22:08:12.5  &  0.42312  &  0.508  &  22.16 	&  c   &  \\  
V162 & 4127  &  11:13:45.68  &  22:04:22.8  &  0.60893  &  0.711  &  22.33 	&  ab  &  -0.007\\  
V163 & 4147  &  11:13:11.63  &  22:10:37.2  &  0.62955  &  0.633  &  22.31 	&  ab  &  -0.010\\  
V164 & 4159  &  11:13:43.33  &  22:11:57.2  &  0.65348  &  0.859  &  22.19 	&  ab  &  -0.061\\  
V165 & 4165  &  11:13:22.33  &  22:06:18.2  &  0.61594  &  0.709  &  22.17 	&  ab  &  -0.012\\  
V166 & 4200  &  11:13:24.30  &  22:11:12.7  &  0.57983  &  0.879  &  22.33 	&  ab  &  -0.012\\  
V167 & 4238  &  11:13:08.79  &  22:11:16.3  &  0.59477  &  0.789  &  22.32 	&  ab  &  -0.009\\  
V168 & 4304  &  11:13:21.62  &  22:07:32.4  &  0.50692  &  1.232  &  22.43 	&  ab  &  -0.009\\  
V169 & 4331  &  11:13:11.79  &  22:11:15.5  &  0.26704  &  0.542  &  22.34 	&  c   &  \\  
\tableline
\end{tabular}
\end{center}

\begin{center}
TABLE II.  DI93 Stars that are non-variable\\
\begin{tabular}{c|c|c} \tableline\tableline
DI93 ID & Star ID & WS Index\\
\tableline
V3 & 4724 & 1.5\\
V4 & 3541 & 2.3\\
V7 & 5486 & 1.1\\
V12 & 1103 & 1.5\\
V13 & 4698 & 1.4\\
V20 & 2365 & 2.8\\
V26 & 3831 & 1.6\\
V27 & 5673 & 2.9\\
V35 & 4612 & 1.2\\
V42 & 4057 & 1.3\\
V45 & 3065 & 0.8\\
V47 & 4421 & 0.9\\
V51 & 5433 & 1.2\\
V58 & 4224 & 1.2\\
V59 & 3939 & 1.2\\
V63 & 3890 & 2.0\\
V66 & 1291 & 1.1\\
V68 & 3558 & 2.7\\
V71 & 4318 & 2.4\\
V72 & 3445 & 2.7\\
\tableline
\end{tabular}
\end{center}

\begin{center}
TABLE III.  Anomalous Cepheids\\
\begin{tabular}{c|c|c|c|c|c|c|c|c} \tableline\tableline
Var ID & Star ID &  RA         &  DEC      & Period  & $A_{V}$ & $m_{V}$  & Pulsation Mode & Swope ID\\
       &         &  J2000.0    & J2000.0   & (Days)  &         &          & &\\
\tableline
V53  & 645  &  11:13:22.39 &  22:08:34.6 & 1.48466 &   1.240 &  20.45 & Fundamental & V27\\
V170 & 737  &  11:13:09.58 &  22:10:26.3 & 1.37955 &   1.050 &  20.59 & Fundemental & V203\\
V171 & 1205 &  11:13:14.54 &  22:11:41.6 & 0.39191 &   0.879 &  21.38 & Overtone & V51\\
V172 & 1545 &  11:13:34.75 &  22:13:44.8 & 0.41907 &   0.850 &  21.70 & Overtone & V1\\
\tableline
\end{tabular}
\end{center}

\begin{figure}
\plotone{Siegel.RRfig3.eps}
\end{figure}
\begin{figure}
\plotone{Siegel.RRfig5.eps}
\end{figure}
\begin{figure}
\plotone{Siegel.RRfig6.eps}
\end{figure}
\begin{figure}
\plotone{Siegel.RRfig7.eps}
\end{figure}
\begin{figure}
\plotone{Siegel.RRfig8.eps}
\end{figure}


\begin{references}
\reference{AOH83} Aaronson, M., Olszewski, E. W. \& Hodge, P. W. 1983, \apj, 267, 271
\reference{AJW83} Azzopardi, M., Lequeux, J. \& Westerlund, B. E. 1985, \aap, 144, 388
\reference{B02} Bailey, S. I. 1902, {\it Annals of Harvard College Observatory}, 38, 1
\reference{BHSZ} Beauchamp, D., Hardy, E., Suntzeff, N. B. \& Zinn, R. 1995, \aj, 109, 1628
\reference{B92} Blanco, V. 1992, \aj, 104, 734
\reference{Blazhko} Blazhko, S. 1907, {\it Astron. Nachr.}, 175, 325
\reference{BCS} Bono, G., Caputo, F. \& Stellingwerf, R.F. 1994, \apj, 423, 294
\reference{BCSM} Bono, G., Caputo, F., Castellani, V. \& Marconi, M. 1997, \aaps, 121, 327.
\reference{BCFPHZ} Buonanno, R., Corsi, C. E., Fusi Pecci, F., Hardy, E. \& Zinn, R. 1985, \aap, 152, 65
\reference{B75} Butler, D. 1975, \apj, 200, 68
\reference{CJS} Carney, B. W., Storm, J. \& Jones, R. V. 1992, \apj, 386, 663
\reference{CG97} Carretta, E. \& Gratton, R. G. 1997, \aaps, 121, 95
\reference{M3} Carretta, E., Cacciari, C., Ferraro, F. R., Fusi Pecci, F. \& Tessicini, G. 1998, \mnras, 298, 1005
\reference{C83} Castellani, V. 1983, MSAIt, 54, 141
\reference{C98} Catelan, M.  1998, \apj, 495, L81 
\reference{CS99} Clement, C. M. \& Shelton, I.  1999, \apjl, 515, L85
\reference{CKH} Cox, A. N., King, D. S. \& Hodson, S. W. 1980, \apj, 236, 219
\reference{CHC} Cox, A. N., Hodson, S. W. \& Clancy, S. P. 1983, \apj, 266, 94
\reference{D84} Da Costa, G. S. 1984, \apj, 285, 483 (D84)
\reference{D88} Da Costa, G. S. 1988, in IAU Symp. 126, The Harlow Shapley Symposium on Globular 
Cluster Systems in Galaxies, eds. J. E. Grindlay \& A. G. D. Philip (Dordrecht: Kluwer), 217 
\reference{DACS} Da Costa, G. S., Armandroff, T. E., Caldwell, N. \& Seitzer, P. 1996, \aj, 112, 2576
\reference{DZLY} Demarque, P., Zinn, R., Lee, Y. W. \& Yi, S. 1999, \apj, {\it accepted}
\reference{DH83} Demers, S. \& Harris, W. E. 1983, \aj, 88, 329 [DH83]
\reference{DI93} Demers, S. \& Irwin, M. J. 1993, \mnras, 261, 657 [DI93]
\reference{D89} Dickens, R. J. 1989, in {\it The Use of Pulsating Stars in Fundamental Problems
in Astronomy}, ed. E. G. Schmidt, (Cambridge: Cambridge University Press), p. 141
\reference{GTO} Gratton, R. G., Tornambe, A. \& Ortolani, S. 1986, \aap, 169, 111
\reference{G97} Grebel, E. K. 1997, RvMA, 10, 29
\reference{G98} Grebel, E. K. 1998, in IAU Symposium 192: The Stellar Content of Local Group Galaxies, eds.
P. Whitelock \& R. Cannon, ASP Conf. Ser. Vol 192, (San Francisco: ASP), p. 1
\reference{GS98} Grebel, E. K. \& Stetson, P. B. 1998, in IAU Symposium 192: The Stellar Content of Local Group Galaxies, eds.
P. Whitelock \& R. Cannon, ASP Conf. Ser. Vol 192, (San Francisco: ASP), p. 11
\reference{GDK} Green, E. M., Demarque, P. \& King, C. R. 1987, The Revised Yale Isochrones and 
Luminosity Functions (Yale University Observatory, New Haven)
\reference{HW} Harrington, R. G. \& Wilson, A. G. 1950, \pasp, 62, 118
\reference{H96} Harris, W.E. 1996, \aj, 112, 1487
\reference{H62} Hodge, P. W. 1962, \aj, 67, 125
\reference{H82} Hodge, P. W. 1982, \aj, 87, 1668
\reference{H95} Holtzman, J. A., Burrows, C. J., Casertano, S., Hester, J. J., Trauger, J. T., Watson, A. 
M. \& Worthey, G.  1995, \pasp, 107, 1065
\reference{HK} Hurley-Keller, D. , Mateo, M. \& Grebel, E. K. 1999, \apj, 523, L25
\reference{K95} Ka{\l}u\`zny, J., Kubiak, M., Szyman\'ski, M., Udalski, A., Krzemin\'ski, W. \& Mateo, M. 1995, 
\aaps, 112, 407 [K95]
\reference{K98} Ka{\l}u\`zny, J., Hilditch, R. W., Clement, C. \& Rucinski, S. M. 1998, \mnras, 296, 347
\reference{LK} Lafler, J. \& Kinman, T. D. 1965, \apjs, 11, 216
\reference{L98} Layden, A. C. 1998, \aj, 115, 193
\reference{LC99a}Lee, J. W.  \& Carney, B. W. 1999a, \aj, 118, 1373
\reference{LC99b} Lee, J. W. \& Carney, B. W. 1999b, \aj, 117, 2868
\reference{LDZ} Lee, Y. W., Demarque, P.  \& Zinn, R.  1990, \apj, 350, 155 [LDZ]
\reference{L91} Lee, Y. W.  1991, \apj, 373, L43
\reference{L95} Lee, M. G. 1995, \aj, 110, 1155 [L95]
\reference{LB} Lynden-Bell, D. 1982, Observatory, 102, 202
\reference{M92} Majewski, S. R., 1992, \apjs, 78, 87
\reference{M99} Majewski, S. R., Siegel, M. H., Patterson, R. J. \& Rood, R. T. 1999, \apj, 520, L33 [M99]
\reference{M98} Mateo, M. 1998, \araa, 36, 435
\reference{MFK} Mateo, M., Fischer, P. \& Krzeminski, W. 1995, \aj, 110, 2166
\reference{MR96} Mighell, K. J. \& Rich, M. R. 1996, \aj, 111, 777 [MR96]
\reference{N85a} Nemec, J. M. 1985a, \aj, 90, 240
\reference{N85b} Nemec, J. M. 1985b, \aj, 90, 204
\reference{N88} Nemec, J. M., Wehlau, A. \& de Oliveira, C.M. 1988, \aj, 96, 528
\reference{N94} Nemec, J. M., Nemec, A. F. L.  \& Lutz, T. E. 1994, \aj, 108, 222
\reference{NB} Norris, J. \& Bessell, M. S. 1978, \apj, 225, L49
\reference{O87} Olszewski, E. W., Aaronson, M.  \& Schommer, R. A. 1987, \aj, 93, 565
\reference{Oo} Oosterhoff, P. Th. 1939, {\it Observatory}, 62, 104
\reference{PMJ} Palma, C., Majewski, S. R. \& Johnston, K. V. 1999, \apj, {\it submitted}
\reference{PSCS} Pritzl, B., Smith, H.A., Catelan, M. \& Sweigart, A. V. 1999, \apj, {\it accepted}
\reference{RMS} Renzini, A., Mengel, J. G. \& Sweigart, A. V. 1977, \aap, 56, 369
\reference{R83} Renzini, A. 1983, MSAIt, 54, 335
\reference{RHS97} Rutledge, G. A., Hesser, J. E. \& Stetson, P. B. 1997, /pasp, 109, 907
\reference{SMS} Saha, A., Monet, D. G. \& Seitzer, P. 1986, \aj, 92, 302
\reference{SKS} Sandage, A., Katem, B. \& Sandage, M. 1981, \apjs, 46, 41 [SKS]
\reference{S81a} Sandage, A. 1981a, \apj, 244, L23
\reference{S81b} Sandage, A. 1981b, \apj, 248, 161
\reference{S82a} Sandage, A. 1982a, \apj, 252, 553
\reference{S82b} Sandage, A. 1982b, \apj, 252, 574
\reference{S93a} Sandage, A. 1993a, \aj, 106, 687 [S93a]
\reference{S93b} Sandage, A. 1993b, \aj, 106, 703
\reference{S94} Sarajedini, A. 1994, \aj, 107, 618
\reference{SD} Smith, H. S. \& Dopita, M. A. 1983, \apj, 271, 113
\reference{Sm95} Smith, H. A. 1995, {\it RR Lyrae Stars} (Cambridge: Cambridge University Press)
\reference{S78} Stellingwerf, R. F. 1978, \apj, 224, 953
\reference{PBS87} Stetson, P. B. 1987, \pasp, 99, 191
\reference{PBS94} Stetson, P. B. 1994, \pasp, 106, 250
\reference{PBS96} Stetson, P. B. 1996, \pasp, 108, 851
\reference{PBS98} Stetson, P. B. 1998, \pasp, 110, 1448
\reference{S86} Suntzeff, N. B., Aaronson, M., Olszewski, E. W. \& Cook, K.H. 1986, \aj, 91, 1091 [S86]
\reference{S91} Suntzeff, N. B., Kinman, T. D. \& Kraft, R. P. 1991, \apj, 367, 528
\reference{S67} Swope, H. H. 1967, \pasp, 79, 439
\reference{S68} Swope, H. H. 1968, \aj, 73, S204
\reference{V73} van Agt, S. 1973, in {\it Variable Stars in Globular
Clusters and Related Systems}, ed. J. D. Fernie, (Dordrecht: Reidel), p. 35
\reference{VB} van Albada, T.S. \& Baker, N. 1971, \apj, 169, 311
\reference{VB85} VandenBerg, D. A. \& Bell, R. 1985, \apjs, 58, 561
\reference{V85} VandenBerg, D. A. 1985, \apjs, 58, 711
\reference{V95} Vogt, S. S., Mateo, M., Olszewski, E. W. \& Keane, M. J. 1995, \aj, 109, 151
\reference{WS} Welch, D. L. \& Stetson, P. B. 1993, \aj, 105, 1813
\reference{ZW} Zinn, R. \& West, M. J. 1984, \apjs, 55, 45
\end{references}
\end{document}